\documentclass[aps,pra,twocolumn,showpacs,floatfix]{revtex4}
\usepackage{epsfig}
\usepackage{graphicx}
\usepackage{dcolumn}
\usepackage{amsthm,amsmath}
\usepackage{pstricks}
\usepackage{color}

\begin{document}

\title{Relativistic extended coupled cluster method for magnetic hyperfine structure constant}
 
\author{Sudip Sasmal$^1$\footnote{sk.sasmal@ncl.res.in}, Himadri Pathak$^1$,
Malaya K. Nayak$^2$, Nayana Vaval$^1$ and Sourav Pal$^1$\footnote{s.pal@ncl.res.in}}

\affiliation{$^1$Electronic Structure Theory Group,
Physical Chemistry Division,
CSIR-National Chemical Laboratory, Pune, 411008, India}

\affiliation{$^2$Theoretical Chemistry Section,
Chemistry Group, Bhabha Atomic Research Centre,
Trombay Mumbai 400085, India}

\begin{abstract}
The article deals with the general implementation of 4-component spinor relativistic
extended coupled cluster (ECC) method to calculate first order property of atoms and
molecules in their open-shell ground state configuration. The implemented relativistic ECC
is employed to calculate hyperfine structure (HFS) constant of alkali metals (Li, Na, K, Rb and Cs),
singly charged alkaline earth metal atoms (Be$^{+}$, Mg$^{+}$, Ca$^{+}$ and Sr$^{+}$)
and molecules (BeH, MgF and CaH). We have compared our ECC results with the calculations
based on restricted active space configuration interaction (RAS-CI) method.
Our results are in better agreement with the available experimental values than those of the RAS-CI values.  
\end{abstract}


\maketitle
\section{Introduction}
The interaction of nuclear moment with the internally generated electromagnetic field by
electrons causes small shift and splitting in the energy levels of atom, molecule or ion.
This interaction is known as hyperfine structure (HFS) \cite{lindgren_book}, which plays a
key role in atomic clock and laser experiments.
A variety of applications including telecommunications, global positioning system,
very-long-baseline interferometry telescopes \cite{vlbi} and test of fundamental concepts
of physics \cite{fund_phys} demand very precise measurement of time, which can be given by an
atomic clock, where the unit of time is defined in terms of frequency at which an atom absorbs
or emits photon during a particular transition.
The laser cooling and atom trapping experiments require the knowledge of HFS as it
influences the optical selection rule and the transfer of momentum from photon to the atom.
In particular, as the line width of transition of laser is much smaller than the
energy difference between two hyperfine labels, the frequency of repumping laser depends
on the separation of hyperfine labels \cite{Phillips_1998}.

The standard model (SM) of particle physics predicts either a zero or a very small
(less than 10$^{-38}$ e.cm) electric dipole moment (EDM) of an electron. Therefore, a
measurable non-zero EDM of an electron can explore the physics beyond SM.
The violation of time reversal (T) or equivalently charge conjugation
(C) and spatial parity (P) symmetry of an atomic/molecular system is responsible for
the non zero EDM of an electron. Unfortunately, the accuracy of the theoretically
estimated P,T -odd interaction constants cannot be mapped with the experiment as there are no
corresponding experimental observables. However, the accuracy of theoretically obtained
P,T -odd interaction constants can be estimated by comparing theoretically obtained HFS
constants with the experimental values as the calculation of both requires an accurate
wave function in the nuclear region and the operator forms are more or less similar.

Hyperfine structure as its name suggest, causes very small shift and splitting in the energy levels
and thus, the treatment of it requires simultaneous inclusion of both relativistic effects and electron correlation
as they are non-additive in nature. The best way to include relativistic effect in
a single determinant theory is to solve the Dirac-Hartree-Fock (DHF) Hamiltonian,
whereas single reference coupled cluster (SRCC) method is known to be
the most efficient to include the dynamic part of the electron correlation \cite{bartlett_2007,pal_1989}.
The SRCC method can be solved either by method of variation or by non-variation.
The non-variational solution of SRCC method is the most familiar, known as normal CC (NCC).
The NCC, being non-variational, does not have the upper bound property of energy.
The generalized Hellmann-Feynman (GHF) theorem and (2n+1) rule,
which states that (2n+1)$^{th}$ order energy derivative can be obtained with
the knowledge upto n$^{th}$ order amplitude derivatives, are not satisfied \cite{monkhorst_1977,bartlett_1984}.
The implication of these theorems save enormous computational effort for the calculation
of higher order properties, which clearly a lack in the NCC.
However, the energy derivatives within the NCC can be obtained by Z-vector approach \cite{zvector_cc} or
Lagrange multiplier method of Helgaker {\it et al} \cite{jorgensen_helgaker}.
However, the GHF theorem and the (2n+1) rule are automatically satisfied in the 
variational CC (VCC).
Among the various VCC methods, expectation value CC (XCC), unitary CC (UCC) and
extended CC (ECC) are the most fimiliar in literature.
The XCC and UCC use Euler type of functional where the left vector is complex conjugate of the right vector.
The detailed discussion on various variational coupled cluster methods can be found in reference \cite{rod_xcc1995}.
The ECC functional proposed by Arponen and coworkers \cite{arponen,arponen_bishop} can
bypass all the problems associated with the Euler type of functional by assuming an energy
functional which deals with the dual space of both right and left vector in a double linked
form. This double linking ensures that the energy and its all order derivatives are size
extensive. As the left and right vectors of the ECC functional are not complex conjugates,
it contains relatively large variational space as compared to corresponding Euler type functional.
The linearized version of ECC, in which the left vector is linear, leads to the equations of NCC \cite{pal_lecc}.
Thus, it can be inferred that ECC wavefunction, which spans more correlated determinantal space
than the NCC, eventually improves the correlation energy as well as energy derivatives.

The manuscript is organized as follows. A brief overview of the ECC method including concise details of magnetic HFS constant are
described in Sec. \ref{sec2}. Computational details are given in Sec. \ref{sec3}. We presented our calculated
results and discuss about those in Sec. \ref{sec4} before making our concluding remark.
We are consistent with atomic unit unless stated.

\section{Theory}\label{sec2}
\subsection{ECC functional}
The ECC functional can be derived by parameterizing both bra and ket states. The parametrization
is done by a double similarity transformation that leads to an alternative approach of many body
problem where the functional is biorthogonal in nature. It is pertinent to note that the double
similarity transformed Hamiltonian is no longer Hermitian as the similarity transformations 
are not unitary. The ECC functional of an arbitrary operator (A) is given by
\begin{eqnarray}
\langle A \rangle = \frac{\langle \Phi_0 | e^{\Sigma^{\prime}}
A e^{\Sigma} | \Phi_0 \rangle}{\langle \Phi_0 | e^{\Sigma^{\prime}} e^{\Sigma}
| \Phi_0 \rangle},
\end{eqnarray}
where $ | \Phi_0 \rangle$ is the DHF reference determinant and $\Sigma^{\prime}$, $\Sigma$
are hole-particle (h-p) destruction and creation operator respectively. Arponen proved
that $\langle \Phi_0 | e^{\Sigma^{\prime}} e^{\Sigma} / \langle \Phi_0 |
e^{\Sigma^{\prime}} e^{\Sigma} |\Phi_0 \rangle$ can be written as $\langle\Phi_0
| e^{\Sigma^{\prime{\prime}}}$, where $\Sigma^{\prime{\prime}}$ is h-p destruction operator.
Therefore, the ECC functional for the operator becomes
\begin{eqnarray}
\langle A \rangle = \langle \Phi_0 | e^{\Sigma^{\prime{\prime}}} e^{-\Sigma} A e^{\Sigma} | \Phi_0 \rangle.
\end{eqnarray}
The diagrammatic structure of $e^{-\Sigma} A e^{\Sigma}$, which can also be written
as $(Ae^{\Sigma})_c $ (where c stands for connected), leads to a terminating series.
However, the diagrams in which $\Sigma^{\prime{\prime}}$ is solely connected to
a single $\Sigma$ leads to disconnected term in the amplitude equation. To avoid
this problem, Arponen has defined two sets of amplitudes, $s$ and $t$, with which the
functional can be written as
\begin{eqnarray}
\langle A \rangle = \langle \Phi_0 | e^{S} (A e^{T})_L | \Phi_0 \rangle_{DL},
\label{derieqn}
\end{eqnarray}
where $L$ means that the $T$ operators right side of $A$ are linked to $A$ vertex
and $DL$ denotes that a $S$ operator must be connected to either $A$ or at least
two $T$ operators. The form of the $S$ and $T$ operators are given by
\begin{eqnarray}
X= \sum\limits_{\stackrel{q_1<q_2\dots}{p_1<p_2\dots}}
t_{p_1p_2 \dots}^{q_1q_2\dots}{a_{q_1}^{\dagger}a_{q_2}^{\dagger} \dots a_{p_2} a_{p_1}} ,
\end{eqnarray}
where X is T when $p$($q$) are hole(particle) index and X is S when $p$($q$) are particle(hole) index.

The analytic energy derivatives can be calculated by using the ECC functional given
in equation \ref{derieqn} where the operator is replaced by a perturbed Hamiltonian. The
field dependent perturbed Hamiltonian is given by
\begin{eqnarray}
H(\lambda) = H+\lambda O = f + v + \lambda O,
\end{eqnarray}
where $H$ is the field independent Hamiltonian, $O$ is an external field and $\lambda$
indicates the strength of the field. $f$ and $v$ are one electron and two electron part
of the field independent Hamiltonian respectively.
Pal and co-workers \cite{pal_ecc_prop} have shown that the ECC analytic derivatives can be
obtained by expanding the ECC functional as a power series of $\lambda$ and making the functional
stationary with respect to cluster amplitudes in progressive orders of $\lambda$.
The zeroth order $k$-body cluster amplitudes, which are sufficient to get the first order
derivative of energy (which is nothing but the expectation value in the light of GHF
theorem), can be obtained by using the following conditions
\begin{eqnarray}
\frac{\delta E^{(0)}}{\delta t_k^{(0)}}=0 , &\,&\,
\frac{\delta E^{(0)}}{\delta s_k^{(0)}}=0 .
\end{eqnarray}
Although ECC functional is a terminating series, the
natural truncation in the single and double model leads to computationally very costly terms.
To avoid the costly terms,
we have used the truncation scheme as proposed by Vaval {\it et al,} \cite{vaval_2013} where
the right exponential of the functional is full within the coupled cluster single and double
(CCSD) approximation and all the higher order double linked terms within the CCSD approximation
are taken in left exponent.
The detailed algebraic expression and diagrammatic of the amplitude equations and first order energy derivative are given 
in Appendix \ref{algebraic} and Appendix \ref{diagrammatic}, including the
nuclear magnetic moment $(\mu)$ and spin quantum number (I) of the atoms (in table \ref{mu} of Appendix \ref{magnetic})
and the experimental bond length of molecules used (in table \ref{bond} of Appendix \ref{magnetic}) in our calculation.
\begin{table}[b]
\caption{ Magnetic hyperfine structure constant (A) of ground state ($^{2}S_{1/2}$) of atoms in MHz }
\begin{ruledtabular}
\begin{center}
\begin{tabular}{lrrrrr}
Atom     &   \multicolumn{2}{c}{This work} & Others & Experiment & $\delta \%$ \\
\cline{2-3}\\
& RAS-CI & ECC  &  &  \\
\hline 
$^{6}$Li & 148.5 & 149.3 & 152.1 \cite{yerokhin_2008} & 152.1 \cite{beckmann_1974} & 1.9 \\
$^{7}$Li & 392.1 & 394.3 & 401.7 \cite{yerokhin_2008} & 401.7 \cite{beckmann_1974} & 1.9 \\
$^{23}$Na & 812.1 & 861.8 & 888.3 \cite{safranova_1999} & 885.8 \cite{beckmann_1974} & 2.8 \\
$^{39}$K & 188.8 & 223.5 & 228.6 \cite{safranova_1999} & 230.8 \cite{beckmann_1974} & 3.3 \\
$^{40}$K & -234.7 & -277.9 & & -285.7 \cite{eisinger_1952} & 2.8 \\
$^{41}$K & 103.6 & 122.7 & & 127.0 \cite{beckmann_1974} & 3.5 \\
$^{85}$Rb & 782.3 & 972.5 & 1011.1 \cite{safranova_1999} & 1011.9 \cite{vanier_1974} & 4.0 \\
$^{87}$Rb & 2651.0 & 3295.7 & & 3417.3 \cite{essen_1961} & 3.7 \\
$^{133}$Cs &  & 2179.1 & 2278.5 \cite{safranova_1999} & 2298.1 \cite{arimondo_1977} & 5.5 \\
$^{9}$Be$^{+}$ & -613.7 & -614.6 & -625.4 \cite{yerokhin_2008} & -625.0 \cite{wineland_1983} & 1.7 \\
$^{25}$Mg$^{+}$ & -568.7 & -581.6 & -593.0 \cite{sur_2005} & -596.2 \cite{itano_1981} & 2.5 \\
$^{43}$Ca$^{+}$ & -733.3 & -794.9 & -805.3 \cite{yu_2004} & -806.4 \cite{arbes_1994} & 1.4 \\
$^{87}$Sr$^{+}$ & -872.1 & -969.9 & -1003.2 \cite{yu_2004} & -1000.5(1.0)\cite{buchinger_1990} & 3.1 \\
\end{tabular}
\end{center}
\end{ruledtabular}
\label{tab3}
\end{table}
\begin{table*}[ht]
\caption{Parallel ($A_{\|}$) and perpendicular ($A_{\perp}$) magnetic hyperfine structure constant of molecules in MHz }
\begin{ruledtabular}
\newcommand{\mc}[3]{\multicolumn{#1}{#2}{#3}}
\begin{center}
\begin{tabular}{lrrrrrrrrrrr}
 &  & \mc{5}{c}{A$_{\|}$} & \mc{5}{c}{A$_{\perp}$}\\
\cline{3-7} \cline{8-12}\\
Molecule & Atom & \mc{3}{c}{This work} & Experiment & $\delta \%$ & \mc{3}{c}{This work} & Experiment & $\delta \%$\\
 \cline{3-5} \cline{8-10}\\
 & & SCF & RAS-CI & ECC & \cite{weltner_book} &  & SCF & RAS-CI & ECC & \cite{weltner_book} & \\
\hline
BeH & $^{1}$H & 84.2 & 177.2 & 204.1 & 201(1) \cite{knight_1972} & 1.5 & 65.8 & 158.7 & 185.6 & 190.8(3) \cite{knight_1972} & 2.8\\
 & $^{9}$Be & -182.7 & -203.3 & -200.6 & -208(1) \cite{knight_1972} & 3.7 & -169.4 & -188.9 & -186.0 & -194.8(3) \cite{knight_1972} & 4.7\\
MgF & $^{19}$F & 168.0 & 255.4 & 320.9 & 331(3) \cite{knight_1971a} & 3.1 & 99.8 & 139.4 & 153.3 & 143(3) \cite{knight_1971a} & 6.7\\
 & $^{25}$Mg & -249.2 & -272.4 & -282.6 &  &  & -239.4 & -260.3 & -270.4 &  & \\
CaH & $^{1}$H & 41.1 & 74.6 & 146.4 & 138(1) \cite{knight_1971b} & 5.7 & 37.5 & 70.9 & 141.9 & 134(1) \cite{knight_1971b} & 5.6\\
 & $^{43}$Ca & -259.5 & -307.9 & -321.6 &  &  & -242.7 & -284.6 & -295.7 &  & 
\end{tabular}
\end{center}
\end{ruledtabular}
\label{molecule_table}
\end{table*}

\subsection{Magnetic hyperfine interaction constant}

The interaction of nuclear magnetic moment with the angular momentum of electrons is
responsible for the magnetic HFS. Thus, it can be viewed as a one body interaction
from the point of view of the electronic structure theory \cite{lindgren_book}.
The magnetic vector potential $(\vec{A})$ at a distance $\vec{r}$ due to a
nucleus $K$ of an atom is 
\begin{equation}
\vec{A}=\frac{\vec{\mu}_k \times \vec{r}}{r^3},
\end{equation}
where $\vec{\mu}_k$ is the magnetic moment of nucleus $K$. The perturbed HFS Hamiltonian
of an atom due to $\vec{A}$ in the Dirac theory is given by
$H_{hyp}= \sum_i^n \alpha_i \cdot \vec{A_i}$, 
where $n$ is the total no of electrons and $\alpha_i$ denotes the Dirac $\alpha$
matrices for the i$^{th}$ electron. Now the magnetic hyperfine constant $(A_J)$
of the $J^{th}$ electronic state of an atom is given by
\begin{eqnarray}
A_J = \frac{1}{IJ} \langle \Psi_J | H_{hyp} | \Psi_J \rangle
     = \frac{\vec{\mu_k}}{IJ} \cdot \langle \Psi_J | \sum_i^n \left( 
\frac{\vec{\alpha}_i \times \vec{r}_i}{r_i^3} \right) | \Psi_J \rangle,
\end{eqnarray}
where $|\Psi_J\rangle$ is the wavefunction of the $J^{th}$ electronic state and
$I$ is the nuclear spin quantum number. For a diatomic molecule, the parallel
($A_{\|}$) and perpendicular ($A_{\perp}$) magnetic hyperfine constant can be
written as
\begin{equation}
A_{\|(\perp)}= \frac{\vec{\mu_k}}{I\Omega} \cdot \langle \Psi_{\Omega} | \sum_i^n
\left( \frac{\vec{\alpha}_i \times \vec{r}_i}{r_i^3} \right)_{z(x/y)} | \Psi_{\Omega(-\Omega)}  \rangle,
\end{equation}
where $\Omega$ represents the z component of the total angular momentum of the diatomic molecule and 
it takes the value of +1/2 in all the considered cases in this manuscript. It is clear from the above 
equation that the $A_{\|}$ is proportional to the diagonal matrix elements of 
$\left( \frac{\vec{\alpha} \times \vec{r}}{r^3} \right)_z$ but $A_{\perp}$ is proportional to the nondiagonal
matrix elements of 
$\left( \frac{\vec{\alpha} \times \vec{r}}{r^3} \right)_{x/y}$
between two different states (+$\Omega$ and -$\Omega$). However, $\Omega$ = +1/2
and -1/2 states are degenerate and their corresponding determinants differ by only one
spin up or spin down electron. Thus, the cluster amplitudes are of same magnitude for both 
$\Omega$ = +1/2 and -1/2 states. So, for each system, cluster amplitudes are evaluated once 
and they are used to calculate both $A_{\|}$ and $A_{\perp}$ with their corresponding
property integrals. However, the rearrangement of the one electron property matrix of
$A_{\perp}$ is necessary in the contraction between individual matrix element and proper
cluster amplitude.

\section{Computational details}\label{sec3}

The DIRAC10 \cite{dirac10} package is used to solve the DC Hamiltonian and to obtain
one-electron hyperfine integrals.
Finite size of nucleus with Gaussian charge distribution is considered as the nuclear model.
The nuclear parameters for the Gaussian charge distribution are taken as default values in DIRAC10.
Aug-cc-pCVQZ basis \cite{aug_ccpcvqz_LiNe, aug_ccpcvqz_NaMg} is used for Li, Be,
Na, Mg, F atoms and aug-cc-pCV5Z \cite{aug_ccpcvqz_LiNe} is used for H 
atom.
We have used dyall.cv4z
\cite{dyall_4s7s} basis for K, Ca and Cs atoms and dyall.cv3z \cite{dyall_4s7s}
basis for Rb and Sr atoms. All the occupied orbitals are taken
in our calculations. The virtual orbitals whose energy exceed a certain
threshold (see in table \ref{ras} of Appendix \ref{magnetic}) are not taken into account in our
calculations as the contribution of high energy virtual orbitals
is negligible in the correlation calculation.
Restricted active space configuration interaction (RAS-CI) calculations are done using
a locally modified version of DIRAC10 package and the detailed description of RAS
configuration is compiled in table \ref{ras} of Appendix \ref{magnetic}.

\section{Results and discussion}\label{sec4}
The numerical results of our calculations of HFS constant using 4-component spinor 
ECC method, capable of treating ground state open-shell configuartion are presented.
We also present results using RAS-CI method. The RAS-CI calculations are done using
DIRAC10 package.

\begin{figure}[b]
\centering
\begin{center}
\includegraphics[height=4.5 cm, width=8.0 cm]{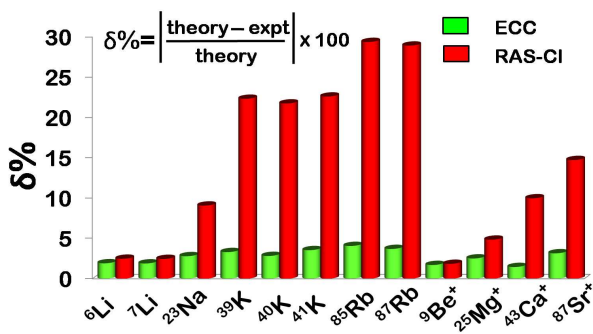}
\caption {Comparison of relative deviations between ECC and RAS-CI values of our calculations.}
\label{figure}
\end{center}  
\end{figure}

In Table \ref{tab3}, we present the HFS constant values of  alkali metal atoms
starting from Li to Cs and singly charged alkaline earth metal atoms (Be$^{+}$
to Sr$^{+}$). Our results are compared with the available experimental
values and the values calculated using RAS-CI method. The deviation of our ECC
values from the experimental values are
presented as $\delta \%$. Our ECC results are in good agreement with the
experimental results ($\delta \%$ $<$ 6$\%$). It is observed that the deviations
increase as we go down both in the alkali metal and in alkaline earth metal group
of the periodic table except for the Ca$^{+}$ ion in the series.
The deviations of RAS-CI and ECC values with the experimental
values are presented in Fig \ref{figure}. 
\begin{table*}[ht]
\caption{HFS constant of $^{1}H$ of CaH molecule in RAS-CI method.}
\begin{ruledtabular}
\begin{center}
{%
\newcommand{\mc}[3]{\multicolumn{#1}{#2}{#3}}
\begin{center}
\begin{tabular}{ccccccccccccc}
\mc{3}{c}{Basis} & \mc{4}{c}{A$_{\|}$ (MHz)} & \mc{4}{c}{A$_{\perp}$ (MHz)}\\
\cline{1-3} \cline{4-7} \cline{8-11} \\
Ca & H & Spinor & SCF & Correlation & Total & Experiment & SCF & Correlation & Total & Experiment\\
\hline 
dyall.v3z & cc-pVTZ & 192 & 38.9 & 35.3 & 74.2 & × & 35.3 & 35.4 & 70.7 & ×\\
dyall.cv3z & aug-cc-pCV5Z & 274 & 41.2 & 33.4 & 74.6 & 138(1) \cite{weltner_book,knight_1971a} & 37.5 & 33.4 & 70.9 & 134(1) \cite{weltner_book,knight_1971a} \\
dyall.cv3z & aug-cc-pCV5Z & 318 & 41.2 & 34.0 & 75.2 & × & 37.5 & 34.1 & 71.6 & ×
\end{tabular}
\end{center}
}%
\end{center}
\end{ruledtabular}
\label{tab6}
\end{table*}
It is clear that
the deviations of RAC-CI are always greater than ECC and it is expected as the
coupled cluster is a better correlated theory than the truncated CI theory.
It is interesting to note that the deviations in RAS-CI increase much faster
rate compared to ECC as we go down the groups.
This reflect the fact that
truncated CI is not size extensive and thus it does not scale properly
with the increasing number of electrons.
It should be noted that the ratio of theoretically estimated HFS constant of
different isotopes must be the ratio of their nuclear g factor for point nuclear model.
Different isotopes are treated by changing the nuclear magnetic moment $(\mu)$ of the atom
but nuclear parameters for each isotopes are same which is by default of the most stable
isotopes in DIRAC10. This causes difference in  $\delta \%$ of different isotopes.

In Table \ref{molecule_table}, we present the parallel and perpendicular HFS constant of ground
state of BeH, MgF and CaH molecules obtained from RAS-CI and ECC theory. We have compared
our ECC results with the available experimental values and the deviations are reported
as $\delta \%$. Our calculated results within the ECC framework show good agreement with the experimental values.
The highest deviation for parallel HFS constant is in the case of $^{1}$H of CaH where
the ECC value differs only $\sim$ 8.5 MHz. This is also better than the RAS-CI values
where the deviation is too off ($\sim$ 63.5 MHz) from the experimental values.
However, for $^{9}$Be in BeH and $^{19}$F in MgF, RAS-CI yields marginally 
better results ($\sim$ 10 MHz) as compared to ECC.

Like the RAS-CI parallel HFS constant values of $^{1}$H of CaH, the perpendicular
HFS constant is very off from the experimental values. To investigate this, we have
calculated the HFS of CaH with more number of virtual orbitals in the same as well as with
a different basis. The results are presented in table \ref{tab6}. It is clear from Table \ref{tab6}
that for this system RAS-CI gives very bad estimation of HFS constant.
A possible explanation is as follows, according to Kutzelnigg's error analysis
\cite{kutzelnigg_1991} the comparative error in CI energy can be written as
[O($\delta$ + O(S$^2$))]$^2$ where $e^S$ is the wave operator of NCC method
and $\delta$ is the error of the wave operator. Although the comparative
error analysis of CI by Kutzelnigg is with respect to NCC but we expect a
similar expression will be hold for ECC also. From Table \ref{tab6}, it is clear
that the DHF (SCF) contribution to the energy derivative is significantly less whereas the
correlation contribution for ECC to the energy derivative is very large as compared
to SCF contribution which is evident from table \ref{molecule_table}.
Therefore, the DHF ground state is very poor reference for
this system and for ECC, the wave operator must be large enough. Thus, it associates
considerably large error in the CI energy as the error in CI energy is proportional to the
quartic of wave operator of CC wave function.

It is interesting to see that both the parallel and perpendicular HFS of $^{1}$H
decrease as we go from BeH to CaH. This indicates that the spin density near
$^{1}$H nucleus of CaH is less than that of BeH. This explains the ionicity
of the bond in CaH is greater than the bond in BeH.

We have done series of calculations to estimate uncertainty in our calculations by comparing our 
ECC results with FCI results taking example of $^7$Li, $^9$Be and  BeH. Details
of the these calculations are included in Appendix \ref{fullci}.
We believe that the uncertainty in our calculations with respect to full CI results for the atomic systems are
well within 5\% and 10\% for the molecular systems considering all possible sources
of error in our calculations.  

\section{Conclusion}\label{sec5}
We have successfully implemented the relativistic ECC method using 4-component 
Dirac spinors to calculate first order energy derivatives of atoms and molecules
in their open-shell ground state configuration. We applied this method to
calculate the magnetic HFS constant of Li, Na, K, Rb, Cs, Be$^{+}$, Mg$^{+}$,
Ca$^{+}$ and Sr$^{+}$ along with parallel and perpendicular magnetic HFS
constant of BaH, MgF and CaH molecules. We also present RAS-CI results to
show the effect of correlation in the calculation of HFS constant.
Our ECC results are in good agreement with the experiment. We have found
some anomalies in RAS-CI results of CaH and given a possible explanation.

\section*{Acknowledgement}
Authors acknowledge a grant from CSIR XII$^{th}$ Five Year
Plan project on Multi-Scale Simulations of Material (MSM) and the resources
of the Center of Excellence in Scientific Computing at CSIR-NCL.
S.S acknowledges the Council of Scientific and Industrial Research (CSIR)
for Shyama Prasad Mukherjee (SPM) fellowship.
\clearpage
\newpage
\onecolumngrid
\appendix
\section{Algebraic expression of ECC energy and cluster amplitude equation}\label{algebraic}
The zeroth order ECC energy functional within the approximation stated in the manuscript is
\begin{equation}
\begin{split}
E^{(0)} =& \langle \Phi_0 | \Big[ vt_2^{(0)} + s_2^{(0)}v + s_1^{(0)}ft_1^{(0)} + s_2^{(0)}ft_2^{(0)}
+ s_1^{(0)}vt_1^{(0)} + s_2^{(0)}vt_2^{(0)} + s_2^{(0)}vt_1^{(0)} + s_1^{(0)}vt_2^{(0)}
+ \frac{1}{2!}vt_1^{(0)}t_1^{(0)}\\
& + \frac{1}{2!}s_1^{(0)}s_1^{(0)}v + \frac{1}{2!}s_1^{(0)}vt_1^{(0)}t_1^{(0)} + \frac{1}{2!}s_1^{(0)}s_1^{(0)}vt_1^{(0)}
+ s_1^{(0)}vt_1^{(0)}t_2^{(0)} + s_2^{(0)}vt_1^{(0)}t_2^{(0)} + \frac{1}{2!}s_2^{(0)}vt_2^{(0)}t_2^{(0)}\\
& + s_1^{(0)}s_2^{(0)}vt_2^{(0)} + \frac{1}{3!}s_2^{(0)}v(t_1^{(0)})^3
+ \frac{1}{2!}s_2^{(0)}vt_2^{(0)}(t_1^{(0)})^2 + \frac{1}{4!}s_2^{(0)}v(t_1^{(0)})^4 
+\frac{1}{3!}(s_1^{(0)})^3vt_2^{(0)} + ft_1^{(0)}\\
& + s_1^{(0)}f + \frac{1}{2!}s_1^{(0)}f(t_1^{(0)})^2  + s_2^{(0)}ft_1^{(0)}t_2^{(0)} + \frac{1}{3!}s_1^{(0)}v(t_1^{(0)})^3 
\Big] | \Phi_0 \rangle.
\end{split}
\end{equation}
Equation for $s_1^{(0)}[\delta E^{(0)} / \delta t_1^{(0)}=0]$ is given by
\begin{equation}
\begin{split}
\langle \Phi_0 | \Big[ & vt_1^{(0)} + s_1^{(0)}(f+v) + s_2^{(0)}v + s_1^{(0)}vt_1^{(0)} + s_1^{(0)}vt_2^{(0)}
+ s_2^{(0)}vt_1^{(0)} + s_2^{(0)}vt_2^{(0)} + \frac{1}{2!}s_1^{(0)}s_1^{(0)}v + s_2^{(0)}vt_2^{(0)}t_1^{(0)}\\
& + \frac{1}{2!}s_2^{(0)}vt_1^{(0)}t_1^{(0)} + \frac{1}{3!}s_2^{(0)}v(t_1^{(0)})^3 + f + s_1^{(0)}ft_1^{(0)}
+ s_2^{(0)}ft_2^{(0)} + \frac{1}{2!}s_1^{(0)}vt_1^{(0)}t_1^{(0)} \Big] | \Phi_{i}^{a} \rangle = 0.
\end{split}
\end{equation}
Equation for $s_2^{(0)}[\delta E^{(0)} / \delta t_2^{(0)}=0]$ is given by
\begin{equation}
\begin{split}
\langle \Phi_0 | \Big[ & v + s_1^{(0)}v + s_2^{(0)}(f+v) + \frac{1}{2!}s_1^{(0)}s_1^{(0)}v + s_1^{(0)}s_2^{(0)}v
+ s_1^{(0)}vt_1^{(0)} + s_2^{(0)}vt_1^{(0)} + s_2^{(0)}vt_2^{(0)} + \frac{1}{2!}s_2^{(0)}vt_1^{(0)}t_1^{(0)} \\
& + \frac{1}{3!}(s_1^{(0)})^3v + s_2^{(0)}ft_1^{(0)} \Big] | \Phi_{ij}^{ab} \rangle = 0.
\end{split}
\end{equation}
Equation for $t_1^{(0)}[\delta E^{(0)} / \delta s_1^{(0)}=0]$ is given by
\begin{equation}
\begin{split}
\langle \Phi_{i}^{a} | \Big[ & (f+v)t_1^{(0)} + vt_2^{(0)} + \frac{1}{2!}vt_1^{(0)}t_1^{(0)} + vt_1^{(0)}t_2^{(0)}
+ s_1^{(0)}v + s_1^{(0)}vt_1^{(0)} + s_1^{(0)}vt_2^{(0)} + s_2^{(0)}vt_2^{(0)}
+ s_1^{(0)}s_1^{(0)}vt_2^{(0)}\\
& + f +\frac{1}{2!}ft_1^{(0)}t_1^{(0)} + \frac{1}{3!}v(t_1^{(0)})^3 \Big] | \Phi_0 \rangle = 0.
\end{split}
\end{equation}
Equation for $t_2^{(0)}[\delta E^{(0)} / \delta s_2^{(0)}=0]$ is given by
\begin{equation}
\begin{split}
\langle \Phi_{ij}^{ab} | \Big[ & v + vt_1^{(0)} + (f+v)t_2^{(0)} + \frac{1}{2!}vt_1^{(0)}t_1^{(0)} + vt_1^{(0)}t_2^{(0)}
+ \frac{1}{2!}vt_2^{(0)}t_2^{(0)} + \frac{1}{2!}vt_2^{(0)}t_1^{(0)}t_1^{(0)} + \frac{1}{3!}v(t_1^{(0)})^3
+ \frac{1}{4!}v(t_1^{(0)})^4 \\ 
& + ft_1^{(0)}t_2^{(0)} + s_1^{(0)}vt_2^{(0)} \Big] | \Phi_0 \rangle = 0.
\end{split}
\end{equation}
Equation for $E^{(1)}$ is given by
\begin{equation}
\begin{split}
E^{(1)} =& \langle \Phi_0 | \Big[ Ot_1^{(0)} + s_1^{(0)}O + s_1^{(0)}Ot_1^{(0)} + s_2^{(0)}Ot_2^{(0)}
+ \frac{1}{2!}s_1^{(0)}Ot_1^{(0)}t_1^{(0)} + s_2^{(0)}Ot_1^{(0)}t_2^{(0)} \Big] | \Phi_0 \rangle.
\end{split}
\end{equation}
\clearpage
\newpage
\section{Diagrammatic of amplitude equation and energy derivative of ECC}\label{diagrammatic}
\begin{figure}[h]
\centering
\begin{center}
\includegraphics[height=8 cm, width=15 cm]{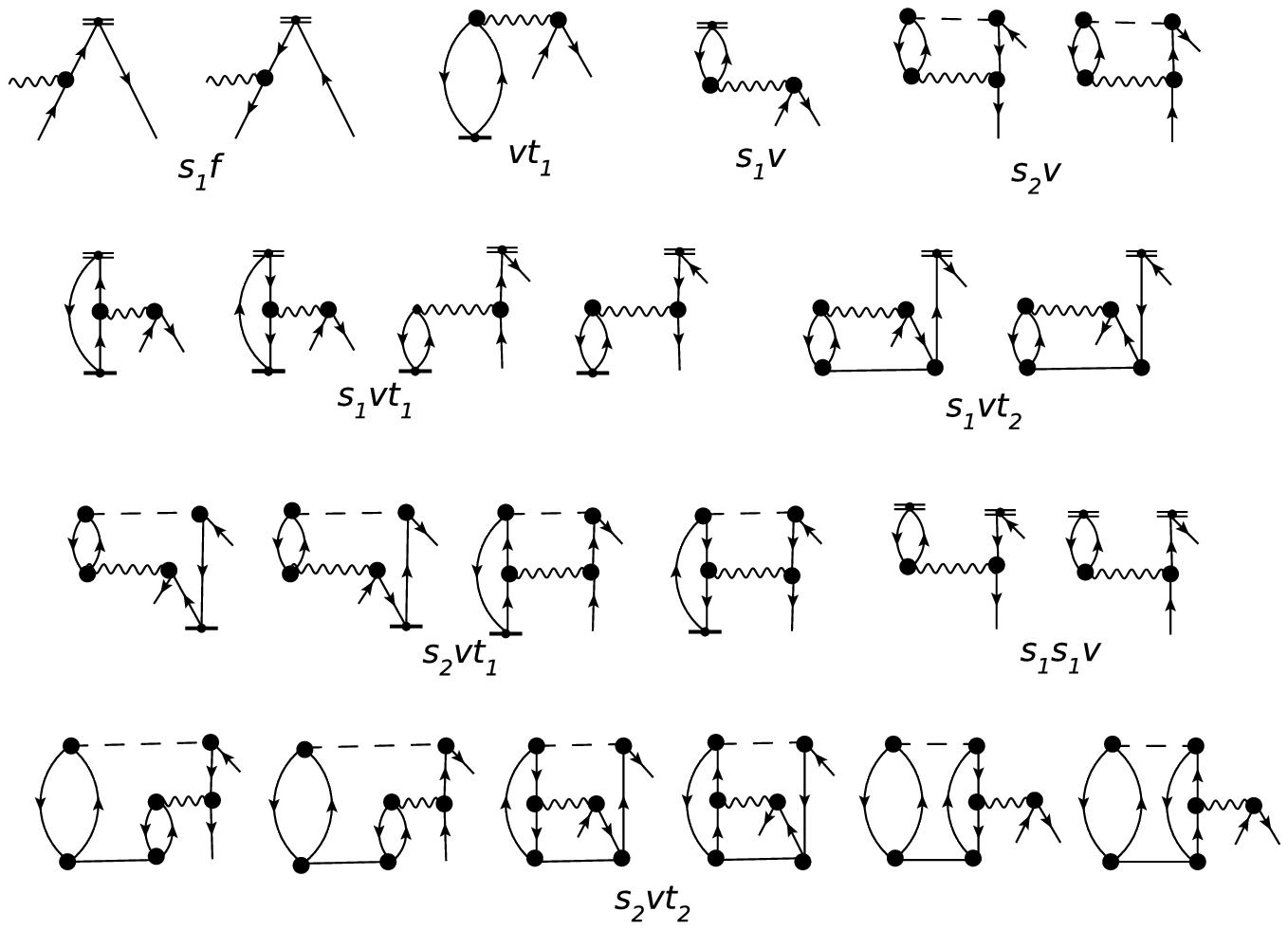}
\includegraphics[height=8 cm, width=15 cm]{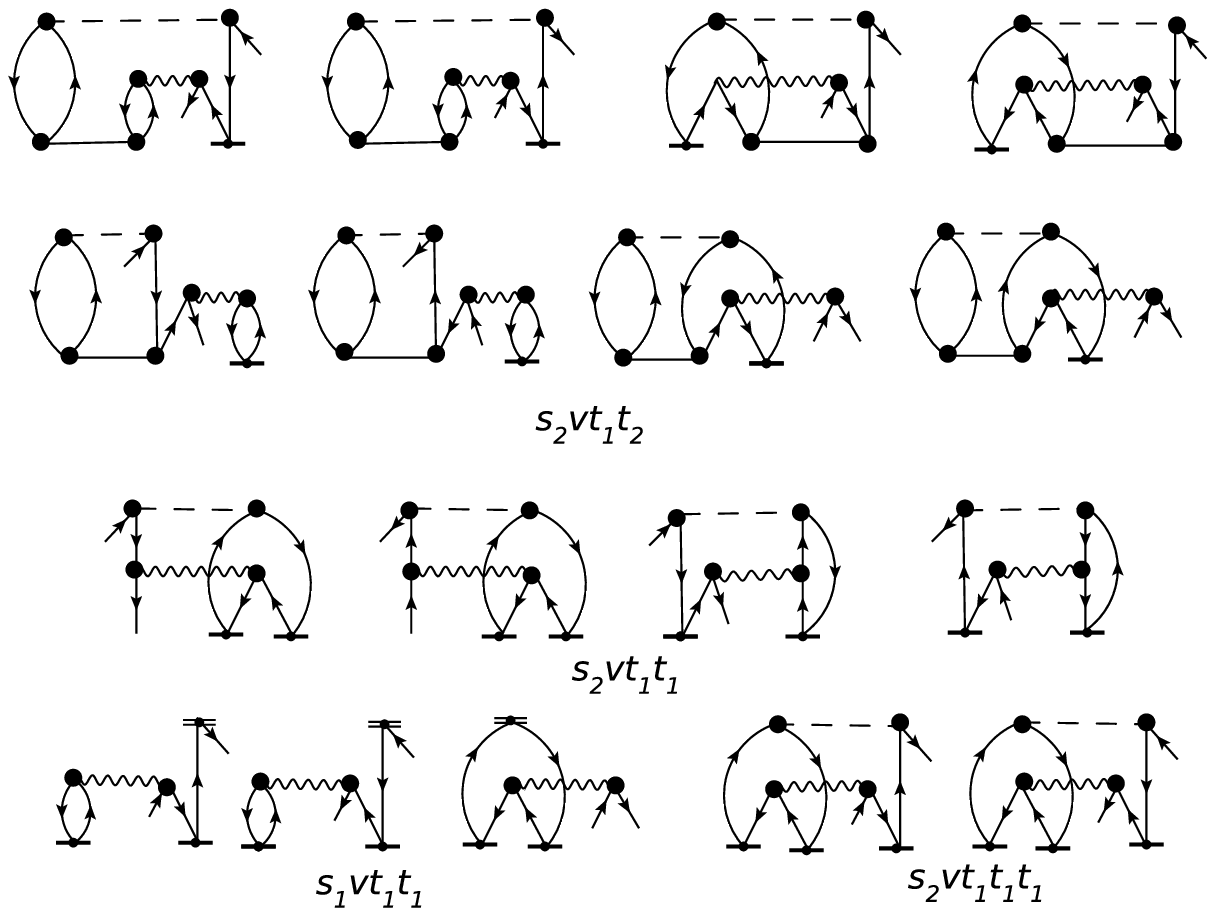}
\includegraphics[height=2 cm, width=9 cm]{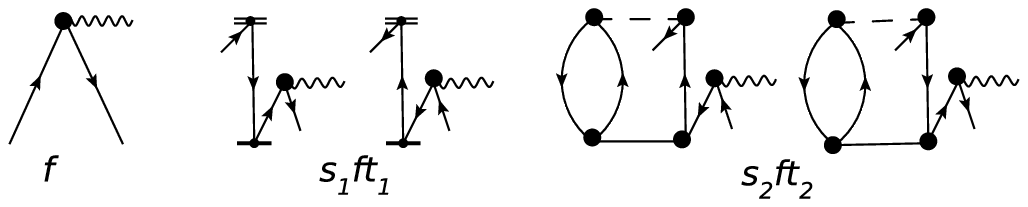}
\caption{{Diagrams for s$_1$ amplitude}}
\label{fig_s1}
\end{center}
\end{figure}

\begin{figure}
\centering
\begin{center}
\includegraphics[height=8 cm, width=15 cm]{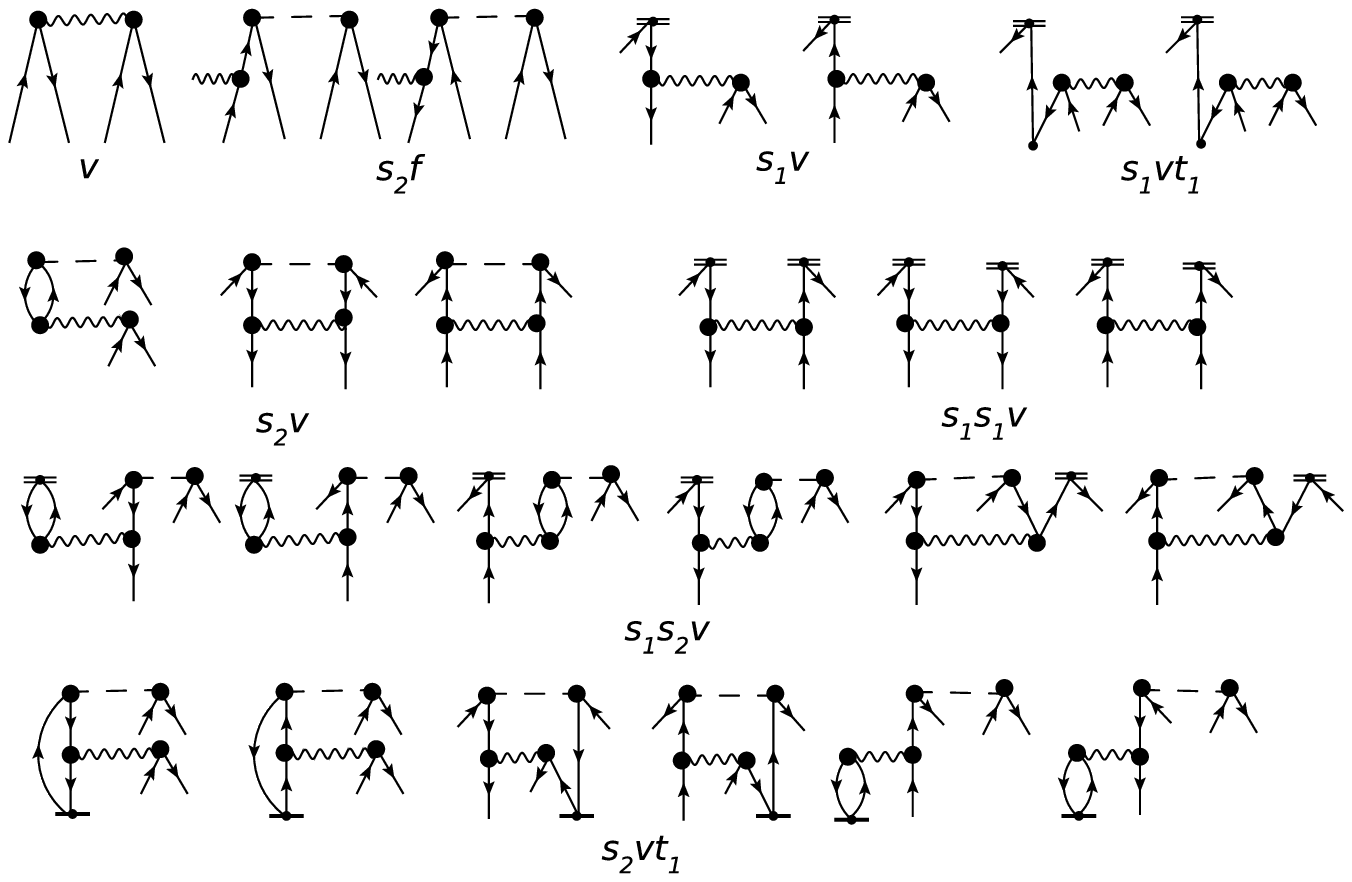}
\includegraphics[height=8 cm, width=15 cm]{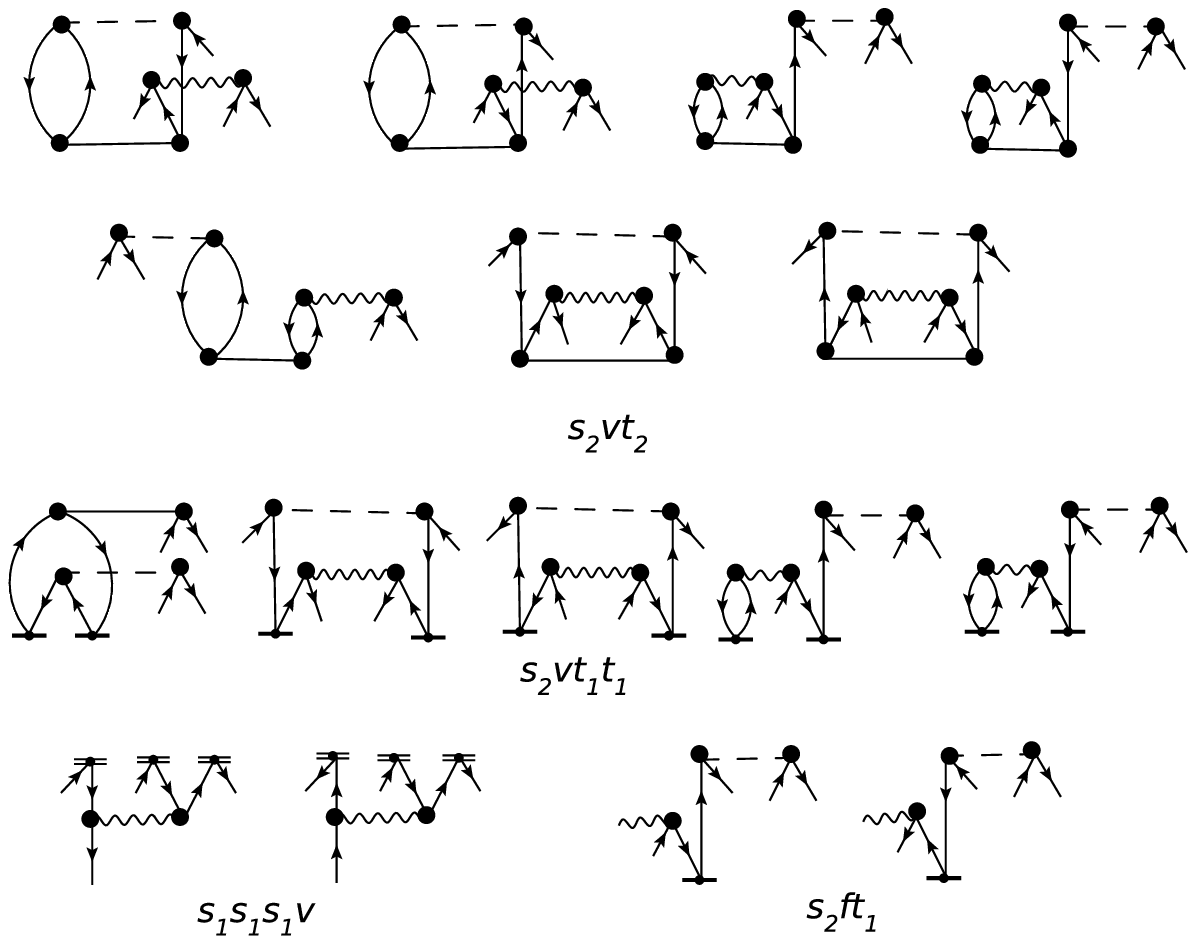}
\caption{{Diagrams for s$_2$ amplitude}}
\label{fig_s2}
\end{center}
\end{figure}

\begin{figure}
\centering
\begin{center}
\includegraphics[height=8 cm, width=15 cm]{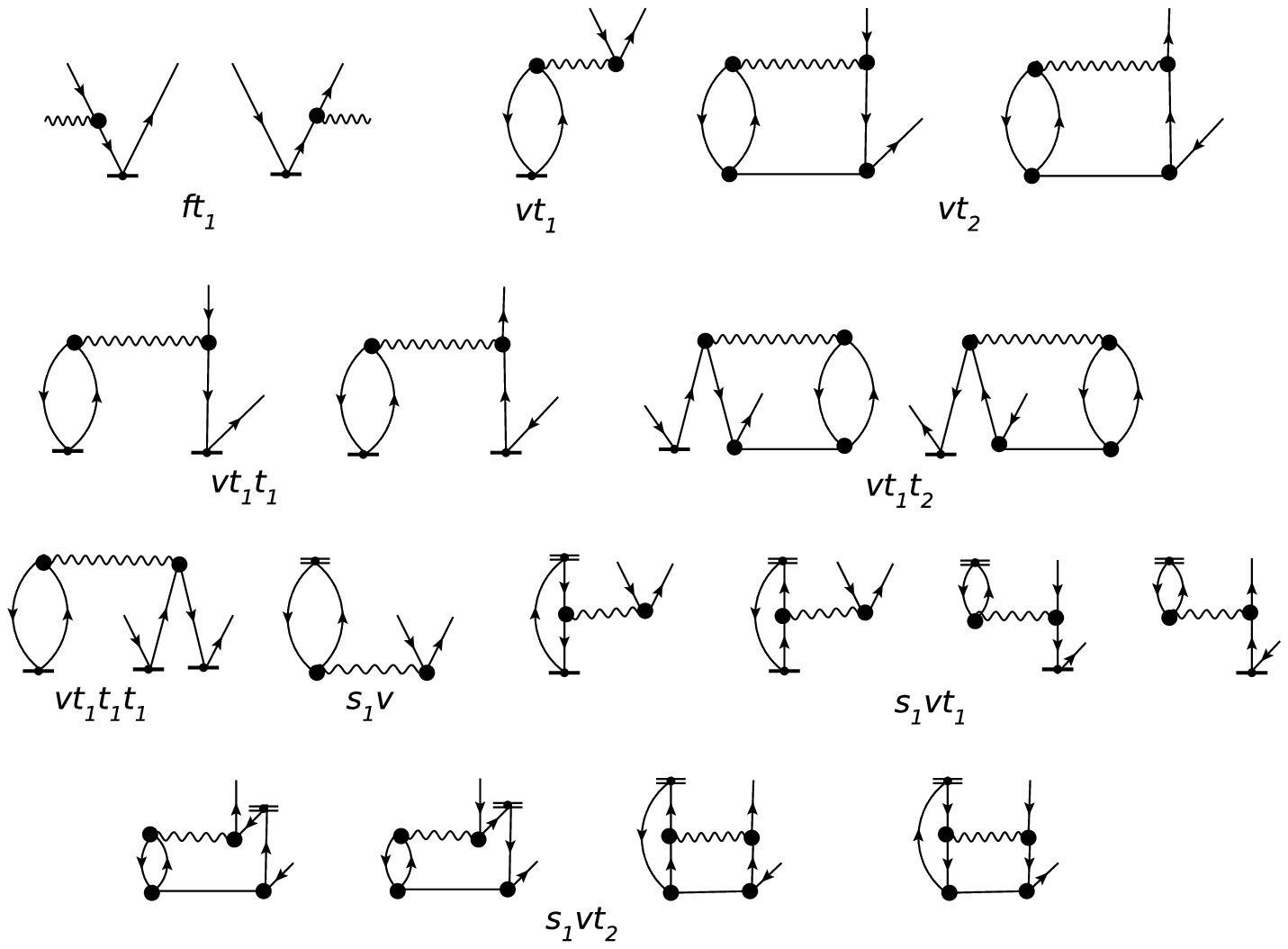}
\includegraphics[height=6 cm, width=15 cm]{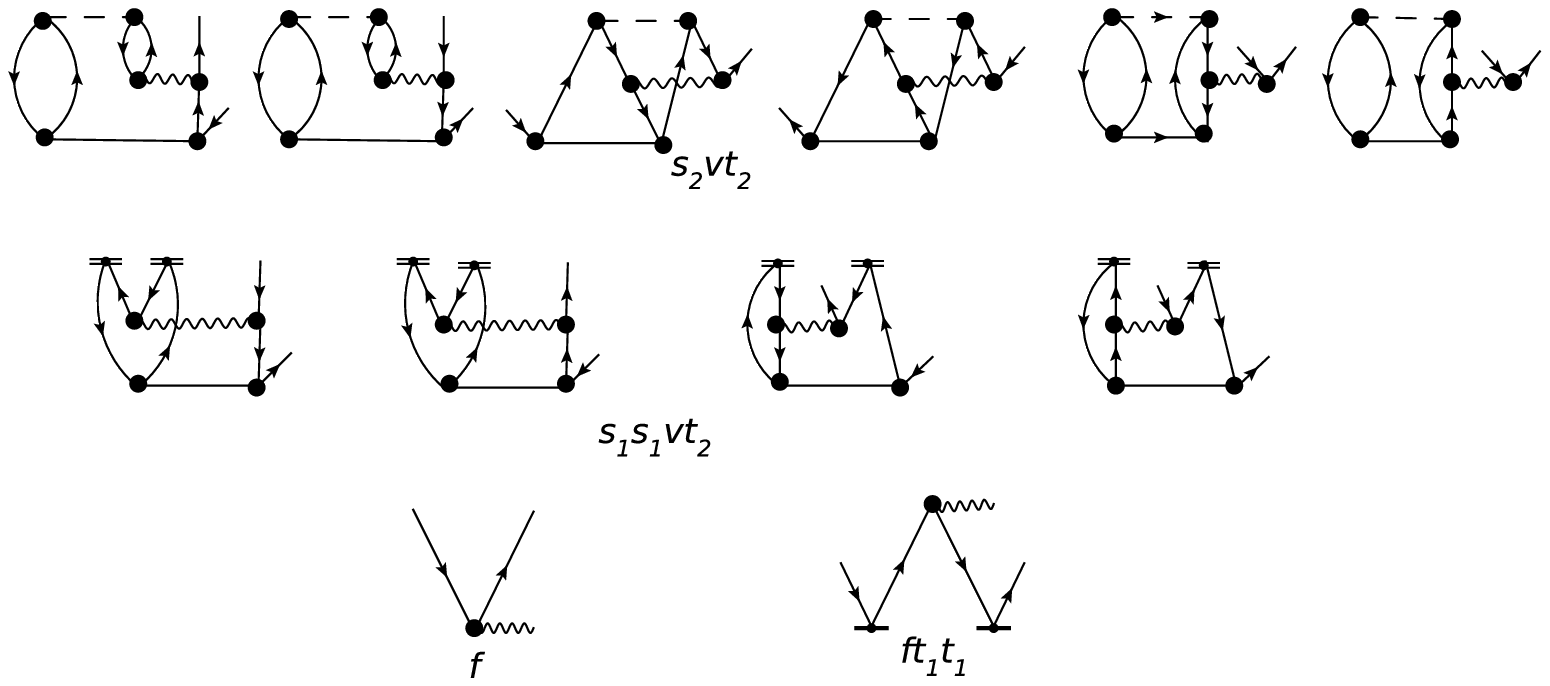}
\caption{{Diagrams for t$_1$ amplitude}}
\label{fig_t1}
\end{center}
\end{figure}

\begin{figure}
\centering
\begin{center}
\includegraphics[height=8 cm, width=15 cm]{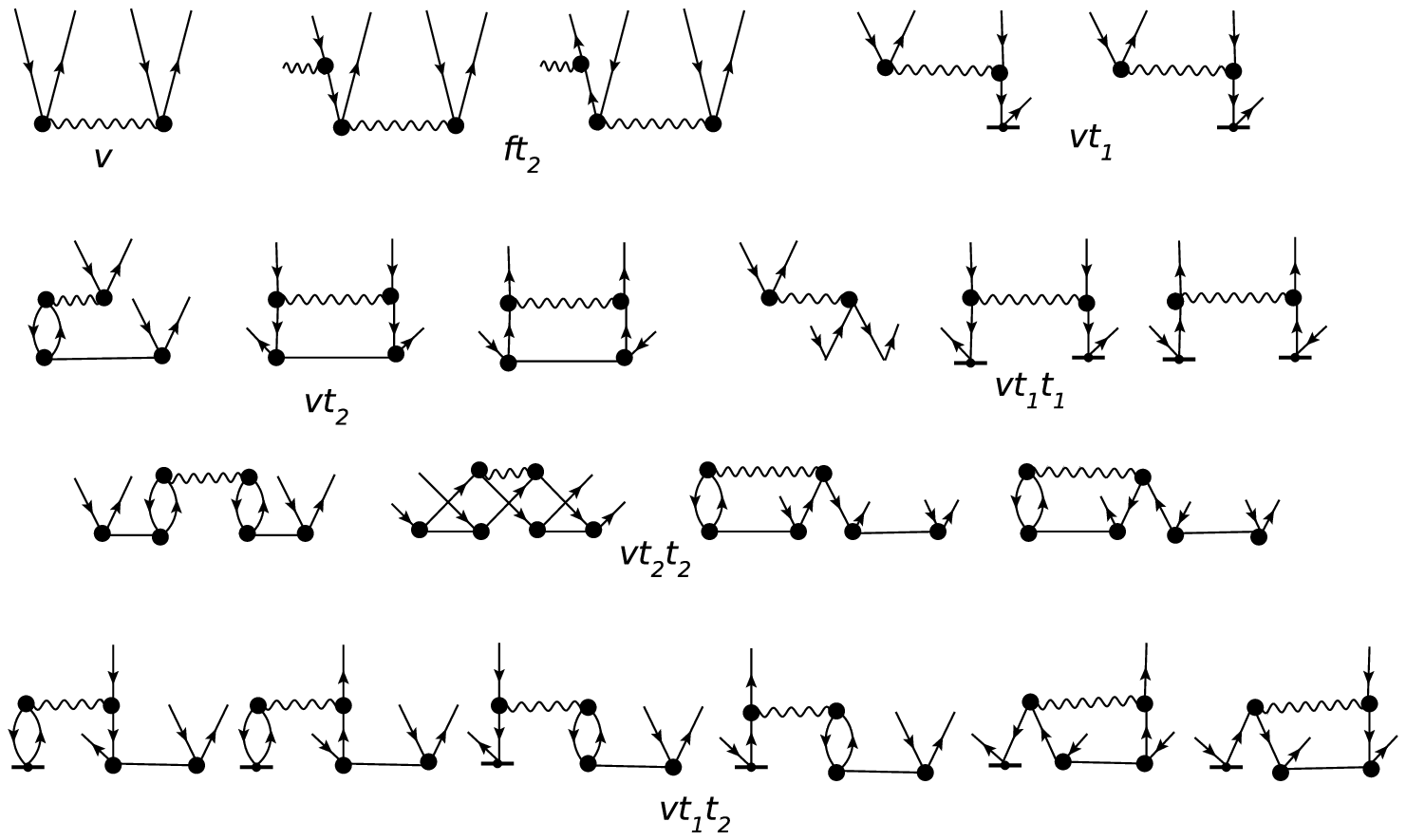}
\includegraphics[height=8 cm, width=17 cm]{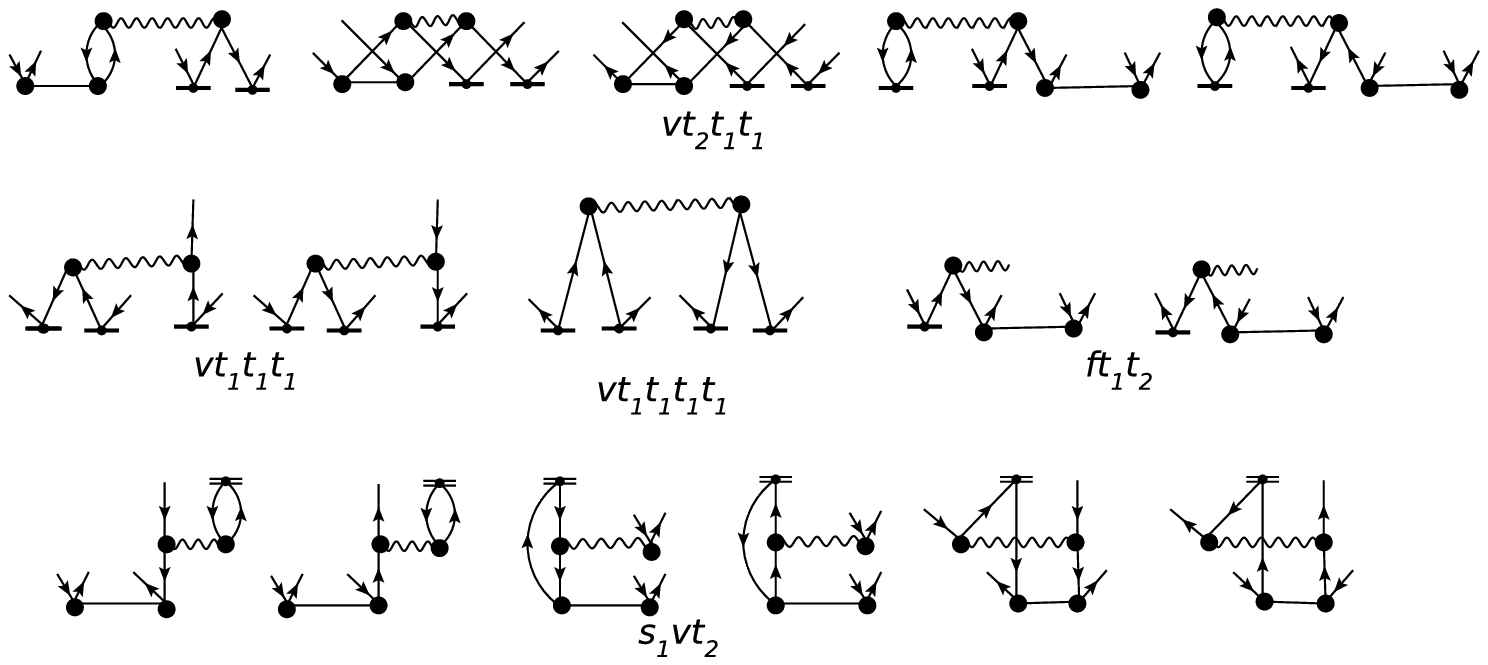}
\caption{{Diagrams for t$_2$ amplitude}}
\label{fig_t2}
\end{center}
\end{figure}

\begin{figure}
\centering
\begin{center}
\includegraphics[height=6 cm, width=13 cm]{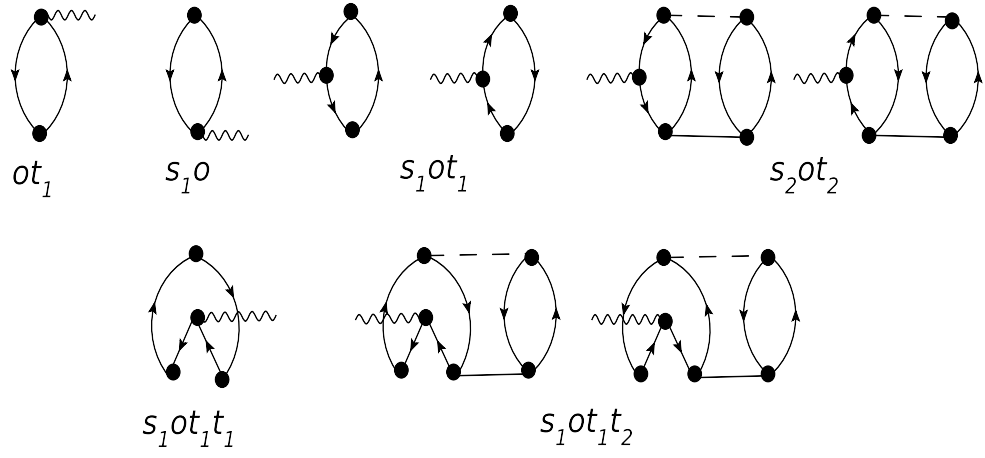}
\caption{{Diagrams for first order energy derivative (E$^{(1)}$)}}
\label{fig_e1}
\end{center}
\end{figure}

In Fig. \ref {fig_s1}, Fig. \ref{fig_s2}, Fig. \ref{fig_t1} and Fig. \ref{fig_t2}, we present all the necessary
diagrams required to construct the equations for s$_1$, s$_2$, t$_1$ and t$_2$ amplitudes respectively.
The diagrams required for first order energy derivative (E$^{(1)}$) are given in Fig. \ref{fig_e1}.
\clearpage
\newpage
\section{Nuclear parameters of the atoms, bond length of the molecules, RAS-CI configuration and threshold energy of atoms and molecules}\label{magnetic}
The nuclear magnetic moment $(\mu)$ and nuclear spin quantum number of atoms used in our calculations
are presented in table \ref{mu}. The experimental bond length for the molecular system are 
presented in table \ref{bond}. RAS-CI configuration and threshold energy for the correlation
calculation of the atomic and molecular system are compiled in table \ref{ras}.
\begin{table*}[h]
\caption{ Nuclear magnetic moment ($\mu$) and nuclear spin quantum no (I) of atoms \cite{mu_I}.}
\begin{ruledtabular}
\begin{center}
\begin{tabular}{rrrrrrrrrrrrrrrr}
Atom&$^{1}$H&$^{2}$D&$^{6}$Li&$^{7}$Li&$^{9}$Be&$^{19}$F&$^{23}$Na& $^{25}$Mg
&$^{39}$K&$^{40}$K&$^{41}$K&$^{43}$Ca&$^{85}$Rb&$^{87}$Rb&$^{87}$Sr \\
I&\,1/2&\,1&\,1&\,3/2&\,3/2&\,1/2&\,3/2&\,5/2&\,3/2&\,4&\,3/2&\,7/2&\,5/2&\,3/2&\,9/2\\
$\mu/\mu_N$&\,2.7928&\,0.8574&\,0.8220&\,3.2564&\,-1.1779&\,2.6288&\,2.2175&\,-0.8554&\,0.3914&\,-1.2981&\,0.2149&\,-1.3172&\,1.3530&\,2.7512&\,-1.0928\\ 
\end{tabular}
\end{center}
\end{ruledtabular}
\label{mu}
\end{table*}
\begin{table}[h]
\caption{ Bond length of the molecules in \AA}
\begin{ruledtabular}
\begin{tabular}{lr}
Molecule & Bond length \cite{bondlen}\\
\hline
{BeH}    &\, 1.343  \\
{MgF}    &\, 1.750 \\
{CaH}    &\, 2.003  \\
\end{tabular}
\end{ruledtabular}
\label{bond}
\end{table}
\begin{table*}[h]
\caption{ RAS-CI configuration and threshold energy of atoms and molecules}
\begin{ruledtabular}
\newcommand{\mc}[3]{\multicolumn{#1}{#2}{#3}}
\begin{center}
\begin{tabular}{lrrr}
Atom/Molecule & \mc{2}{c}{RAS Configuration$^{a}$} & Threshold energy$^{b}$ \\
\cline{2-3}
× & RAS I & RAS II & (a.u.)\\
\hline
Li & 2, 1 & 3, 4 & $\infty$ \\
Na & 6, 5 & 3, 4 & $\infty$ \\
K & 10, 9 & 3, 4 & 500 \\
Rb & 19, 18 & 3, 4 & 500 \\
Cs & × & × & 60 \\
Be$^+$ & 2, 1 & 3, 4 & $\infty$ \\
Mg$^+$ & 6, 5 & 3, 4 & $\infty$ \\
Ca$^+$ & 10, 9 & 3, 4 & 500 \\
Sr$^+$ & 19, 18 & 3, 4 & 100 \\
BeH & 3, 2 & 3, 4 & $\infty$ \\
MgF & 11, 10 & 3, 4 & 10 \\
CaH & 11, 10 & 5, 6 & 15 \\
\end{tabular}
\end{center}
\end{ruledtabular}
$^{a}$ In each RAS configuration spin up and spin down spinors are seperated by comma. \\
Maximum number of holes in RAS I is 2. Maximum number of electrons in RAS III is 2. \\
$^{b}$ $\infty$ value means all the spinors are considered in the correlation calculation.
\label{ras}
\end{table*}
\clearpage
\newpage
\section{Comparison of full CI and ECC HFS constant values}\label{fullci}
The comparison of full CI and ECC HFS constant values of $^{7}$Li and $^{9}$Be$^{+}$ is
presented in table \ref{li} and table \ref{be} respectively. The comparison of parallel and 
perpendicular component of full CI and ECC HFS constant values of BeH is compiled in
table \ref{beh}.

\begin{table}[h]
\caption{ Comparison of full CI and ECC HFS values (in MHz) of $^{7}$Li}
\begin{ruledtabular}
\begin{tabular}{lrr}
Basis & Full CI & ECC\\
\hline
aug-cc-pCVDZ &\, 384.1  &\, 383.9 \\
aug-cc-pCVTZ &\, 402.0 &\, 401.5 \\
aug-cc-pCVQZ$^{a}$ &\, 386.0 &\, 385.5
\end{tabular}
\end{ruledtabular}
\label{li}
$^{a}$ Considering 3 electrons and 189 virtual orbitals
\end{table}
\begin{table}[h]
\caption{ Comparison of full CI and ECC HFS values (in MHz) of $^{9}$Be$^{+}$}
\begin{ruledtabular}
\begin{tabular}{lrr}
Basis & Full CI & ECC\\
\hline
aug-cc-pCVDZ &\, -586.6  &\, -586.5 \\
aug-cc-pCVTZ &\, -615.7 &\, -615.6 \\
aug-cc-pCVQZ$^{a}$ &\, -613.0 &\, -612.8
\end{tabular}
\end{ruledtabular}
\label{be}
$^{a}$ Considering 3 electrons and 183 virtual orbitals
\end{table}
\begin{table}[h]
\caption{ Comparison of full CI and ECC HFS values (in MHz) of BeH.}
\begin{ruledtabular}
\newcommand{\mc}[3]{\multicolumn{#1}{#2}{#3}}
\begin{center}
\begin{tabular}{lrrrrr}
Basis & Atom & \mc{2}{c}{$A_{\|}$} & \mc{2}{c}{$A_{\perp}$}\\
\cline{3-4} \cline{5-6} 
× & × & Full CI & ECC & Full CI & ECC\\
\hline
cc-pVDZ & $^{9}$Be & -158.7 & -159.3 & -145.9 & -146.6\\
× & $^{1}$H & 189.9 & 187.6 & 174.5 & 172.2\\
aug-cc-pVDZ & $^{9}$Be & -165.5 & -166.1 & -152.6 & -153.2\\
× & $^{1}$H & 188.7 & 186.2 & 172.1 & 169.7
\end{tabular}
\end{center}
\end{ruledtabular}
\label{beh}
\end{table}


\newpage
\clearpage
\twocolumngrid

\end{document}